\documentclass[preprint,nofootinbib,superscriptaddress,pacs]{revtex4}
\usepackage{graphicx}
\usepackage{amssymb}
\usepackage{color}
\begin{document}

\title{Neutron skin thickness of $^{48}$Ca, $^{132}$Sn and $^{208}$Pb with KIDS density functional}
\author{Chang Ho Hyun}
\affiliation{Department of Physics Education, Daegu University, Gyeongsan 38453, Korea}

\date{\today}

\begin{abstract}
Neutron skin thickness ($R_{np}$) of the atomic nuclei $^{48}$Ca, $^{132}$Sn and $^{208}$Pb are considered in the 
KIDS (Korea:IBS-Daegu-Sungkyunkwan) density functional formalism.
Model parameters most relevant to $R_{np}$ are constrained from standard nuclear data and neutron star properties
determined from modern observations.
Mean values of $R_{np}$ are obtained as 0.161, 0.220 and 0.160~fm for $^{48}$Ca, $^{132}$Sn and $^{208}$Pb, respectively,
and the corresponding standard deviations are 0.011, 0.020 and 0.021~fm.
Correlations between $R_{np}$ of the three nuclei are considered, and we find that $R_{np}$'s are correlated very strongly.
We obtain correlation coefficient 0.978 between $R_{np}$ of $^{48}$Ca and $^{208}$Pb,
and 0.996 between $R_{np}$ of $^{132}$Sn and $^{208}$Pb.
\end{abstract}

\maketitle

{\it Introduction}

Equation of state (EoS) of infinite nuclear matter is typically characterized by the parameters
$K_0$, $J$, $L$, and $K_{\rm sym}$ over a range of density accessible in laboratory experiments 
and observation of compact objects in the universe. 
The parameters are defined in the energy per nucleon in nulcear matter as
\begin{eqnarray}
E(\rho,\, \delta) &=& E(\rho) + S(\rho) \delta^2 + O(\delta^4), \nonumber \\
E(\rho) &=& E_0 + \frac{1}{2} K_0 x^2 + \cdots , \nonumber \\
S(\rho) &=& J + L x + \frac{1}{2}K_{\rm sym} x^2 + \cdots,
\end{eqnarray}
where $x = (\rho - \rho_0)/ 3\rho$ and $\delta = (\rho_n - \rho_p)/\rho$.
$\rho_n$, $\rho_p$ are the neutron and proton density, respectively,
$\rho$ is the baryon density ($= \rho_n + \rho_p$ in the nucleon matter),
and $\rho_0$ is the nuclear saturation density.
It is frequent in nature to have $\delta \neq 0$ at densities where the role of $K_0$, $L$ and $K_{\rm sym}$ becomes important.
Neutron skin thickness of nuclei, and the neutron star EoS are representative examples that highlight the role of $L$ and $K_{\rm sym}$.
For the correct description of nuclear structure at $\rho < \rho_0$ and the EoS of nuclear matter at $\rho > \rho_0$
where $\delta$ is substantially away from 0, precise determination of the EoS parameters $K_0$, $J$, $L$, $K_{\rm sym}$
is an unavoidable requirement.

In a recent work \cite{ksym},  as an attempt to reduce the uncertainties of the EoS parameters,
the parameter space ($K_0$, $J$, $L$, $K_{\rm sym}$) was surveyed over the region admitted in the literature
$230 \leq K_0 \leq 260$, $30 \leq J \leq 34$, $40 \leq L \leq 70$, $-420 \leq K_\tau \leq -240$
where the units are in MeV, and $K_\tau$ is defined as $K_\tau = K_{\rm sym} - (6 + Q_0/K_0) L$.
Dividing the interval of each parameter by 5, 0.5, 5 and 20 MeV for $K_0$, $J$, $L$, and $K_\tau$ respectively,
the number of EoS considered initially is 4410.
Each energy density functional (EDF) is transformed to a Skyrme-type force.
Skyrme force has two constants that are absent in the nuclear matter EDF: the coefficients of $\nabla^2$ and spin-orbit terms.
The two constants are fit to 13 nuclear data: the energy and the radius of $^{16}$O, $^{40}$Ca, $^{48}$Ca, $^{90}$Zr, $^{132}$Sn,
and $^{208}$Pb, and the energy of $^{218}$U.
With the 4410 Skyrme force models fit to 13 nuclear data, in order to quantify the accuracy of a model,
we calculate the average deviation per datum (ADPD) 
\begin{equation}
{\rm ADPD}(N) = \frac{1}{N} \sum^N_{i=1} \left| \frac{O^{\rm exp}_i - O^{\rm cal}_i}{O^{\rm exp}_i} \right|,
\end{equation} 
where $O^{\rm exp}_i$ ($O^{\rm cal}_i$) denotes an observable from experiment (theory).
In order to reduce the model space, we assume ADPD $< 0.3$~\% and obtain 358 models satisfying the condition.
We call the set of 358 models ADPD03.
The model space is reduced further by imposing a constraint on the radius of $1.4M_\odot$ neutron star 
$R_{1.4} = 11.8 - 12.5$~km.
Among the 358 models in ADPD03, 158 models fulfill the constraint of $R_{1.4}$. We call the set of 158 models R14.

In this work we investigate the neutron skin thickness $R_{np}$ of $^{48}$Ca, $^{132}$Sn and $^{208}$Pb with ADPD03 and R14 sets.
For $^{48}$Ca and $^{208}$Pb, accurate measurements have been performed in recent experiments.
PREX Collaboration reported the first stage measurement $R^{208}_{np}$(PREX-I) = 0.15 $-$ 0.49 fm \cite{prex1}.
The accuracy has been improved in the upgraded second stage experiment $R^{208}_{np}$(PREX-II) = 0.212 $-$ 0.354 fm.
Indirect determination is also accessible by measuring the $E1$ polarizability $\alpha_D$ and pygmy dipole resonance.
Measurement of $\alpha_D$ at RCNP reports $R^{208}_{np}(\alpha_D) = 0.135-0.181$ fm \cite{alphaD208},
and pygmy resonance experiment gives $R^{208}_{np}({\rm pygmy}) = 0.20 - 0.28$ fm \cite{pygmy208}.
Measurement of the $\alpha_D$ has also been applied to $^{48}$Ca, and the result is obtained as
$R^{48}_{np}(\alpha_D) = 0.14 - 0.20$ fm \cite{alphaD48}.
Recent experiment at RIKEN performed the scattering of $^{48}$Ca on the C target, and extracted a result
$R^{48}_{np}({\rm RIKEN}) = 0.146 \pm 0.048$ fm \cite{riken}.
Measurement of $R^{48}_{np}$ in the parity-violating electron scattering was performed at JLab.
CREX Collaboration announced ``thin skin" of $^{48}$Ca in the press release, and the numerical results are expected 
to be available soon. 
It's been shown that $R^{132}_{np}$ and $R^{208}_{np}$ are almost linearly correlated \cite{prc2012}.
Confirming the correlation is important for checking the consistency of theory predictions and experimental results.
Measurement of $R^{132}_{np}$ in the R3B collaboration also strongly motivates the calculation of $R^{132}_{np}$.

{\it Result}

\begin{table}
\begin{center}
\begin{tabular}{l|cc|cc}\hline
 & \multicolumn{2}{c|}{ADPD03} & \multicolumn{2}{c}{R14} \\ \hline
 & mean & s.d. & mean & s.d. \\ \hline
$K_0$ & 252.0 & 5.6 & 251.1 &  5.3 \\
$J$ & 30.9 & 0.7 & 30.7 & 0.6 \\ 
$L$ & 54.7 & 9.7 & 49.8 & 5.2 \\
$K_{\rm sym}$ & -52.4 & 72.0 & -82.4 & 33.7 \\ 
$\mu_S$ & 0.98  & 0.09& 0.97 & 0.09  \\
$\mu_V$ & 0.81 & 0.07 & 0.80 & 0.07 \\
\hline
\end{tabular}
\end{center}
\caption{Mean and standard deviation (s.d.) of EoS parameters $K_0$, $J$, $L$, $K_{\rm sym}$ in the unit of MeV,
and the isoscalar and isovector effective masses to the nucleon mass in free state $\mu_S$ and $\mu_V$
from the 358 models in the ADPD03 set and the 158 models in the R14 set.}
\label{tab1}
\end{table}

Table \ref{tab1} summarizes the mean and the standard deviationof the EoS parameters $K_0$, $J$, $L$, $K_{\rm sym}$,
and the isoscalar and isovector effective mass ratios to the free nucleon mass at the saturation density $\mu_S = m^*_S/m_N$ and 
$\mu_V \ m^*_V/m_N$.
The results of $K_0$ and $J$ are similar between ADPD03 and R14 sets in both mean and standard deviation.
It's been shown in \cite{ksym} that the correlation of $K_0$ and $J$ with $R_{1.4}$ is 0.24 and 0.22 respectively in the extended R14 (eR14) set.
\footnote{In \cite{ksym}, round off is applied to $R_{1.4}$, so the actual range is $11.75 \leq R_{1.4} \leq 12.49$~km.
In this work, we don't use round off, so the models are selected for $11.80 \leq R_{1.4}\leq 12.50$~km.
Therefore, the number of models in the eR14 set is slightly larger than the R14 set.} 
This weak correlation might be an origin of small difference in the mean and the standard deviation of $K_0$ and $J$ between
ADPD03 and R14.
The result also implies that $K_0$ and $J$ could be determined accurately from a proper application of the theory to standard nuclear properties.

Contrary to $K_0$ and $J$, $L$ and $K_{\rm sym}$ exhibit substantial relocation of the mean value and reduction of the standard
deviation when the $R_{1.4}$ constraint is applied.
This dramatic change of $L$ and $K_{\rm sym}$ from the ADPD03 set to the R14 set could also be understood in terms of the correlation with $R_{1.4}$.
Correlation coefficients of $L$ and $K_{\rm sym}$ with $R_{1.4}$ are 0.79 and 0.90 in the eR14 sets, respectively,
so significant changes happen in $L$ and $K_{\rm sym}$ when the constraint of $R_{1.4}$ is taken into account.
 
Mean values and the ranges of $J$ and $L$ of the R14 set are consistent with the result in \cite{lim2013}, $J=29.0 - 32.7$~MeV, and $L=40.5-61.9$~MeV.
Result of $K_{\rm sym}$ in the R14 set is also consistent with the results in the literature (see \cite{ksym} for detail),
and suggests reduced range of the uncertainty.

Within the KIDS formalism, $\mu_S$ and $\mu_V$ can be assumed to take specific values without affecting the nuclear matter
properties at the saturation density, and the basic properties of nuclei \cite{prc2019}.
In this work, however, we don't assume specific values for $\mu_S$ and $\mu_V$, and they are obtained as results of fitting to 13 nuclear data.
The result in Tab.~\ref{tab1} shows that $\mu_S$ and $\mu_V$ are weakly dependent on the model set.
In recent works \cite{eAex, eAin}, $\mu_S$ and $\mu_V$ values are probed in the electron-nucleus scattering in the quasielastic region.
Considered values are $(\mu_S,\, \mu_V) =$ (1.0, 0.8), (0.7, 0.7), (0.9, 0.9), where the first set is obtained without assuming 
specific values of $\mu_S$ and $\mu_V$.
In the result of the electron-nucleus scattering, when the dependence on the effective mass is negligible, three effective mass sets give
results similar to each other, and agree well with data.
In the case where the dependence on the effective mass is clear which corresponds to the response function 
in the exclusive scattering of $^{16}$O, and the cross section in the inclusive scattering of $^{12}$C and $^{40}$Ca,
theory shows better agreement to data with $\mu_S$ close to 1.
Dependence on the symmetry energy was also explored in \cite{eAin}, but the result turns out to be weakly affected by $J$, $L$ and $K_{\rm sym}$.
On the average, models in ADPD03 and R14 have the effective mass consistent with the quasielastic electron scattering data.

\begin{table}
\begin{center}
\begin{tabular}{c|cccc|cccc}\hline
 & \multicolumn{4}{c|}{ADPD03} & \multicolumn{4}{c}{R14} \\ \hline
 & mean &  s.d. &min & max 
 & mean &  s.d. & min & max \\ \hline
$R^{48}_{np}$   & 0.162 & 0.011 & 0.134 & 0.187 & 0.161 & 0.011 & 0.134 & 0.182  \\
$R^{132}_{np}$ & 0.224 & 0.018 & 0.175 & 0.267 & 0.220 & 0.020 & 0.175 & 0.259  \\
$R^{208}_{np}$ & 0.164 & 0.020 & 0.112 & 0.212 & 0.160 & 0.021 & 0.112 & 0.203  \\ \hline
\end{tabular}
\end{center}
\caption{Mean, standard deviation, minimum and maximum values of the neutron thickness of 
$^{48}$Ca, $^{132}$Sn and $^{208}$Pb obtained from the KIDS-ADPD03 and KIDS-R14 models.}
\label{tab2}
\end{table}

In Tab.~\ref{tab2} we summarize the results of the neutron skin thickness of $^{48}$Ca, $^{132}$Sn, and $^{208}$Pb
from the ADPD03 and the R14 sets.
Mean values and standard deviations of the R14 set are similar to those of the ADPD03 set.
The similarity is surprising and unexpected because the $L$ value, known as a key parameter to determine $R_{np}$,
of the R14 set is non-negligibly different from that of the ADPD03 set.
Most notable difference is that the maximum values are slightly reduced in the R14 set,
but the minimum values in the ADPD03 set are unchanged in the R14 set.
Correlation coefficient of $R^{208}_{np}$ and $R_{1.4}$ is $-0.02$ in the ADPD03 set and 0.09 in the eR14 set \cite{ksym},
so the determination of $R_{np}$ is seldom influenced by inclusion or omission of the $R_{1.4}$ constraint.
The result demonstrates that such an insensitivity is not limited to $^{208}$Pb, but valid globally from light to heavy nuclei.

Compared to recent measurement, the result of $R^{48}_{np}$ is compatible with the result of RCNP ($0.14 - 0.20$~fm),
and RIKEN ($0.098 - 0.194$~fm).
In \cite{zhang2021}, $R^{48}_{np}$ is evaluated by using 18 experimental data.
The result of evaluation is $R^{48}_{np} = 0.178 - 0.204$~fm, which is within the range of RCNP and RIKEN,
but much above the result of this work.
As for $R^{208}_{np}$, the mean values of ADPD03 and R14 sets are compatible with PREX-I ($0.15-0.48$~fm)
and $\alpha_D$ ($0.135-0.181$~fm).
However, our result is much lower compared to PREX-II ($0.212-0.354$~fm) and pygmy resonance ($0.20-0.28$~fm).
While the PREX-I result covers almost all the experimental ranges, there is no overlapping region between $\alpha_D$
and PREX-II or pygmy resonance.
Work \cite{zhang2021} also performed the evaluation of $R^{208}_{np}$ with 20 experimental data, and obtained a range
$R^{208}_{np} = 0.156 -0.178$~fm.
The result agrees to our result, and supports small values of $R^{208}_{np}$.

Since the result of CREX experiment is expected to be announced in near future, it is worthwhile to refer to theoretical predictions 
of $R^{48}_{np}$ from various theories. We summarize the result of three works.
In Ref. \cite{prc2012}, $\Delta R_{np}$ of $^{48}$Ca, $^{132}$Sn, and $^{208}$Pb are calculated from the correlation 
between $\alpha_D$ and $R_{np}$ of $^{208}$Pb using 48 reasonable nuclear energy density functionals.
Among the 48 models, sorted out are 25 models that satisfy the data of $\alpha_D$ at RCNP.
Resulting 25 models give $0.176 \pm 0.018$ fm, $0.232 \pm 0.022$ fm and $0.168 \pm 0.022$ fm for 
$R^{48}_{np}$, $R^{132}_{np}$, and $R_{np}^{208}$, respectively.
Center value and uncertainty of $R^{48}_{np}$ are noticeably larger than our result, so though an overlapping region exists 
between the two results, general trend is divided clearly.
On the other hand, our results for the mean values of $R^{132}_{np}$ and $R^{208}_{np}$ are smaller than the center values in \cite{prc2012},
but large portion of the ranges is shared by our result and that of \cite{prc2012}.

Another interesting theory work is the coupled-cluster calculation with the nuclear potential obtained from low-energy
effective field theory (EFT) \cite{nat2016}. It is well known that EFT is a qualified tool for the description of few nucleon systems.
The method is extended to the calculation of $R^{48}_{np}$, and the result is obtained as $R^{48}_{np} = 0.12-0.15$~fm.
The upper most part of the range is consistent with our result, but there is no overalp with the result of \cite{prc2012}.

In \cite{xu2020}, ranges of $J$ and $L$ are inferred from the Bayesian method using the data of $R_{np}$ of Sn isotopes and the neutron star properties.
It is concluded that $R^{48}_{np}$ needs to be larger than 0.15~fm and smaller than 0.25~fm to be compatible with the Sn and/or neutron star data.
Uncertainty is broad, but the work suggests a lower limit of $R^{48}_{np}$.

\begin{table}
\begin{center}
\begin{tabular}{c|ccccc||cccc}\hline
 & $C_{K_0}$ & $C_J$ & $C_L$ & $C_{K_{\rm sym}}$ & $C_{208}$ 
 & lin. app. & PREX-I & PREX-II & $\alpha_D$  \\ \hline
$R^{48}_{np}$ & -0.626 & 0.577 & 0.645 & -0.551 & 0.978 & $0.524x + 0.077$ & 0.16 -- 0.33 & 0.188 -- 0.262 & 0.148 -- 0.172\\
$R^{132}_{np}$ & -0.605 & 0.627 & 0.712 & -0.468 & 0.996 &  $0.976x + 0.064$ & 0.21 -- 0.54 & 0.271 -- 0.410 & 0.196 -- 0.241 \\
$R^{208}_{np}$ & -0.590 & 0.611 & 0.698 & -0.458 & 1      & $x$ & 0.15 -- 0.49 & 0.212 -- 0.354 & 0.135 -- 0.181 \\ \hline
\end{tabular}
\end{center}
\caption{Left section: Correlation coefficients $C_i$ for $i = K_0$, $J$, $L$, $K_{\rm sym}$ and $R^{208}_{np}$
from the R14 set. Right section: Linear approximation of $R^{48}_{np}$ and $R^{132}_{np}$ as a function of $x=R^{208}_{np}$, 
and their ranges obtained by subsitituing the data of $R^{208}_{np}$ in the linear function.}
\label{tab3}
\end{table}

Correlation between the EoS parameters, between the EoS parameters and observables, and between the observables 
have been considered extensively in many works.
In the left section of Tab.~\ref{tab3}, we show the correlation of $R_{np}$ with the EoS parameters and $R^{208}_{np}$ for the R14 set.
Slope of the symmetry energy $L$ shows the strongest correlation with $R_{np}$ among the EoS parameters.
The result is consistent with the observations in the literature that $L$ is most strongly correlated with $R_{np}$.
Quantitative magnitude however does not show overwhelmingly strong correlation compared to other EoS parameters.
It's been argued that the correlation of $L$ with nuclear properties is stronger at $\rho \simeq \frac{2}{3} \rho_0$ than $\rho = \rho_0$.
It may be a reason for $C_L$ not very close to 1.
EoS parameter most uncorrelated with $R_{np}$ is $K_{\rm sym}$.
It's been discussed at Tab.~\ref{tab1} that $K_0$ and $J$ can be determined accurately with the basic nuclear properties.
Therefore accurate measurements of $R_{np}$ will provide a unique chance to determine $L$ unambiguously.

\begin{figure}
\begin{center}
\includegraphics[width=6in]{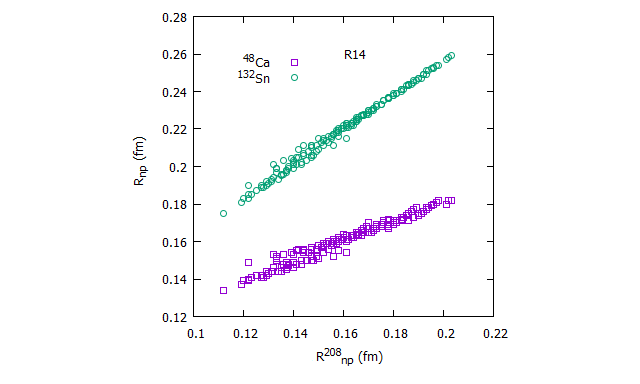}
\end{center}
\caption{Distribution of $R^{48}_{np}$ and $R^{132}_{np}$ with respect to $R^{208}_{np}$. Correlations are,
as shown in Tab.~\ref{tab3}, 0.978 and 0.996, respectively.}
\label{fig1}
\end{figure}

In \cite{prc2012}, the correlations of $R^{48}_{np}$ and $R^{132}_{np}$ with $R^{208}_{np}$ are considered with 48 frequently used EDFs.
Results are 0.852 for $R^{48}_{np}$ and 0.997 for $R^{132}_{np}$.
Extended model space is surveyed in \cite{tagami2020},
and the correlation between $R^{48}_{np}$ and $R^{208}_{np}$ is obtained 0.99 from 206 EoSs.
In order to visualize the correlation of this work, we plot $R^{48}_{np}$ and $R^{132}_{np}$ as functions of $R^{208}_{np}$
for the R14 set in Fig.~\ref{fig1}.
Distribution of $R^{132}_{np}$ is almost a straigh tline.
The correlation coefficient is 0.996, similar to the value of \cite{prc2012}, so it indicates that the distribution is really very close to a straight line.
Distribution of $R^{48}_{np}$ is, compared to $R^{132}_{np}$, more scattered, but the correlation coefficient is 0.978, so it is also close to a straight line.

In the right section of Tab.~\ref{tab3}, linear approximation of $R^{48}_{np}$ and $R^{132}_{np}$ are given as functions of $R^{208}_{np}$
in the column `lin. app.'.
Following columns show the ranges of $R^{48}_{np}$ and $R^{132}_{np}$ obtained by substituting the data of $R^{208}_{np}$
in the linear functions.
Information from the correlation is simple: If one $R_{np}$ is thin, other $R_{np}$'s are also thin, 
and if one $R_{np}$ is thick, other $R_{np}$'s are thick too.
Correlation coefficients obtained in \cite{prc2012}, \cite{tagami2020} and this work are similar to each other,
so the strong correlation between $R_{np}$'s is not limited to specific nuclear models, 
but seems to be a general feature of the nuclear structure.
If the results of $R_{np}$'s obtained from other theories or experiments are not consistent with the correlation,
it casts a challenging issue to the nuclear structure theory.


{\it Summary}

We calculated the neutron skin thickness of $^{48}$Ca, $^{132}$Sn and $^{208}$Pb
with the models constrained by accurate reproduction of nuclear data and the radius of $1.4 M_\odot$ neutron stars
determined from astronomical observation.
The set of models determined from the nuclear data constitutes 385 combinations of the EoS parameters ($K_0$, $J$, $L$, $K_{\rm sym}$),
and the neutron star radius constraint picks 158 models out of the 358 models.
Mean and standard deviation of $R^{48}_{np}$, $R^{132}_{np}$, and $R^{208}_{np}$ are not different much between the two sets,
and they are consistent with results of other theories.
Result of $R^{208}_{np}$ is consistent with range determined from the electric dipole polarizability experiment,
but there is no overlap with the result of PREX-II experiment.

Correlations between $R^{48}_{np}$, $R^{132}_{np}$ and $R^{208}_{np}$ are calculated with 158 models, and they turn out to be very close to 1.
From the linear approximation of $R^{48}_{np}$ and $R^{132}_{np}$ with respect to $R^{208}_{np}$, we obtained ranges of
$R^{48}_{np}$ and $R^{132}_{np}$ corresponding to the data of $R^{208}_{np}$ from PREX-I, PREX-II and electric dipole polarizability.
Ranges corresponding to PREX-II are incompatible with the ranges obtained from the electric dipole polarizability data.
Result of $R^{48}_{np}$ in the CREX collaboration, and measurement of $R^{132}_{np}$ in the R3B collaboration are quite expected.

\section*{Acknowledgments}
This work was supported by the National Research Foundation of Korea (NRF) grant
funded by the Korea govenment (No. 2018R1A5A1025563 and No. 2020R1F1A1052495).

\end{document}